\documentclass[twocolumn,superscriptaddress,preprintnumbers,a4paper,amsmath,amssymb,showpacs,floatfix]
{revtex4-1}
\usepackage{graphicx}
\usepackage{graphics}
\usepackage{bm}
\usepackage{dcolumn}

\newcommand{\urrs}{U(Ru$_{1-x}$Rh$_{x}$)$_{2}$Si$_{2}$}
\newcommand{\urxrs}{U(Ru$_{0.92}$Rh$_{0.08}$)$_{2}$Si$_{2}$}
\newcommand{\urs}{URu$_{2}$Si$_{2}$}
\newcommand{\tcs}{ThCr$_{2}$Si$_{2}$}

\begin{document}

\title{Magnetic structure in \urxrs~single crystal studied by neutron diffraction in static magnetic fields up to 24 T}

\author{K.~Proke\v{s}}
\affiliation{Helmholtz-Zentrum Berlin f\"{u}r Materialien und Energie, Hahn-Meitner Platz 1, 14109 Berlin, Germany}

\author{M. Bartkowiak}
\affiliation{Helmholtz-Zentrum Berlin f\"{u}r Materialien und Energie, Hahn-Meitner Platz 1, 14109 Berlin, Germany}

\author{O. Rivin}
\affiliation{Helmholtz-Zentrum
Berlin f\"{u}r Materialien und Energie, Hahn-Meitner Platz 1, 14109 Berlin, Germany}
\affiliation{Physics Department, Nuclear Research Centre-Negev, P.O. Box 9001, Beer-Sheva 84190, Israel}

\author{O. Prokhnenko}
\affiliation{Helmholtz-Zentrum
Berlin f\"{u}r Materialien und Energie,  Hahn-Meitner Platz 1, 14109 Berlin, Germany}

\author{T. F\"orster}
\affiliation{Hochfeld-Magnetlabor Dresden (HLD-EMFL), Helmholtz-Zentrum Dresden-Rossendorf, D-01314 Dresden, Germany}

\author{S. Gerischer}
\affiliation{Helmholtz-Zentrum
Berlin f\"{u}r Materialien und Energie,  Hahn-Meitner Platz 1, 14109 Berlin, Germany}

\author{R. Wahle}
\affiliation{Helmholtz-Zentrum
Berlin f\"{u}r Materialien und Energie, Hahn-Meitner Platz 1, 14109 Berlin, Germany}

\author{Y.-K. Huang}
\affiliation{Van der Waals-Zeeman Institute, University of Amsterdam, 1018XE Amsterdam, The Netherlands}

\author{J. A.~Mydosh}
\affiliation{Kamerlingh Onnes Laboratory and Institute Lorentz, Leiden University, 2300 RA Leiden, The Netherlands }

\begin{abstract}
We report the high-field induced magnetic phase in single crystal of \urxrs. Our neutron study combined with high-field magnetization, shows that the magnetic phase above the first metamagnetic transition at $\mu_{0}$H$_{c1}$ = 21.6 T has an uncompensated commensurate antiferromagnetic structure with propagation vector \textbf{\textit{Q$_ {2}$}} = ($\frac{2}{3}$ 0 0) possessing two  single-\textbf{\textit{Q}} domains. U moments of 1.45 (9) $\mu_{B}$ directed along the $c$ axis are arranged in an up-up-down sequence propagating along the $a$ axis, in agreement with bulk measurements. The U magnetic form factor at high fields is consistent with both the U$^{3+}$ and U$^{4+}$ type. The low field short-range order that emerges from the pure \urs~due to Rh-doping is initially strengthened by the field but disappears in the field-induced phase. The tetragonal symmetry is preserved across the transition but the $a$ axis lattice parameter increases already at low fields. Our results are in agreement with itinerant electron model with 5$f$ states forming bands pinned in the vicinity of the Fermi surface that is significantly reconstructed by the applied magnetic field.
\end{abstract}

\maketitle

Despite more than three decades of intense study the ground state of the well-established hidden order (HO at T$_{HO}$ = 17.5~K) / superconducting (T$_{sc}$ = 1.5~K) heavy-fermion system \urs~remains unknown and under heavy dispute \cite{urrshfn6,*urrshfn7}. The HO is linked to an antiferromagnetic (AF) order \cite{urrshfn1,urrshfn2,urrshfn3} and fluctuations characterized by a propagation vector \textbf{\textit{Q$_ {0}$}} = (1 0 0) that can be stabilized either by strain or doping. ~\cite{urrshfn37} At temperatures where the HO exists, new phases can be created by perturbations. Strong magnetic field is necessary to suppress the HO order and drive the system into distinct metamagnetic transitions (MT) between 35 and 39 T before reaching a polarized Fermi-liquid state \cite{urrshfn5,urrshfn8}. 

High critical fields can be reduced by a suitable light doping, in particular by Rh substitution for Ru \cite{urrshfn4,urrshfn5}. Such substitutions quickly destroy both the HO and SC states in the range of a few percent, keeping the heavy-fermion behavior intact and stabilize (2 - 3 \% Rh) AF order with \textbf{\textit{Q$_ {0}$}} \cite{urrshfn32}. For doping levels above~10~\% Rh a long-range AF order with \textbf{\textit{Q$_ {3}$}} = ($\frac{1}{2}$ $\frac{1}{2}$ $\frac{1}{2}$) appears \cite{urrshfn4}.

Although high field bulk measurements disclose important insights regarding physical states emerging from HO, only microscopic methods as neutron diffraction yield information regarding the periodicity and nature of the order. However, it is most challenging to combine this technique with static magnetic fields exceeding  $\sim$ 17 T. This is due to two main limitations. On one hand, the magnetic field strengths are limited by the construction material of the magnet, on the other, the orientation of the sample with respect to the magnetic field and neutron beam imposes specific geometrical restrictions.

Challenging neutron experiments in pulsed magnetic fields~\cite{urrshfn9} were recently performed revealing the main features of field-induced magnetic structures in pure \urs~and 4 \% Rh doped systems  that are found to be different \cite{urrshfn12,urrshfn10}. While in the former system it is spin-density wave (SDW) characterized by \textbf{\textit{Q$_ {1}$}} = (0.6 0 0), suspected to be of multi-\textbf{\textit{Q}} nature~\cite{urrshfn12}, in the latter case the propagation vector is \textbf{\textit{Q$_ {2}$}} = ($\frac{2}{3}$ 0 0). However, these experiments suffer from a reduced signal to noise ratio because of the pulsed field nature of the magnetic field and inability to survey a large portion of the reciprocal space during a single field pulse due the use of a triple-axis spectrometer that prevent the detector tilt out of the scattering plane. 

Here we report results achieved at recently completed HFM-EXED facility at Helmholtz-Zentrum Berlin (HZB) that enables neutron studies in static fields up to 26 T - a study of the field-induced magnetic phase in \urxrs, in which we deliberately suppressed HO and allowed for short-range magnetic order. 

The details of our \urxrs~single crystal preparation, quality and other physical properties are presented elsewhere \cite{urrshfn16}. For the present neutron studies a 4$\times$4$\times$4 mm$^{3}$ single crystal was cut using spark erosion to have edges along principal axes and glued onto a copper holder which was placed in the cryostat capable of reaching 1.4 K. This was inserted into a high field magnet HFM - a hybrid solenoid (13 T, 4 MW resistive insert and series-connected 13 T superconducting outsert)~\cite{urrshfn13,urrshfn18}. HFM is situated at a time-of-flight spectrometer (EXED) \cite{urrshfn14}. Further facility details are given in Supplemental Material \cite{urrshfn31}. 

Reduced critical field enabled us to study the first field-induced phase of \urxrs~in static fields in two orientations. Initially, the tetragonal axis was directed along the static field enabling an instantaneous detection of eight nuclear and four of field-induced magnetic Bragg reflections indexed by two, in tetragonal symmetry equivalent, field-induced propagation vectors eliminating drawbacks of pulsed experiments. In the second experiment, in order to observe the signal connected with short-range order at \textbf{\textit{Q$_ {3}$}} and nuclear reflection 101 in forward detector, we rotated the $c$ axis by about 21$^{\circ}$ in the horizontal plane. Here, we were able to record three nuclear and four magnetic reflections. Due to the strong uniaxial response of \urxrs, this angular deviation does not change the physics of our system \cite{urrshfn20} and leads only to lower effective field acting along the $c$ axis. Details about the diffraction geometry are provided in Supplemental Material \cite{urrshfn31}.  

In contrast to the forward direction that is sensitive to both magnetic and nuclear Bragg reflections predominantly only nuclear reflections are observable in the backscatter detector panels. The detected eleven nuclear reflections (two of them, 200 and 101 in the forward direction) allowed us to verify the crystal structure in zero and elevated fields (of the \tcs~type, shown the inset of Fig.~\ref{fig1}), to establish the lattice constants and orientation matrix of the crystal with respect to the laboratory system and to evaluate the magnetic moment magnitudes. We found that the intensities recorded in zero field agree well after necessary absorption and extinction corrections with the crystal structure parameters determined previously \cite{urrshfn16}.

Magnetization $M$($T$) measurements in fields up to 58~T generated by discharging a capacitor bank producing a 25~ms long pulse were performed on a 12.5 mg single crystal originating from the same parent piece. The magnetic signal was detected using compensated pick-up coils and scaled to match values obtained in static fields up to 14 T \cite{urrshfn16}.

\begin{figure}
\includegraphics*[scale=0.22]{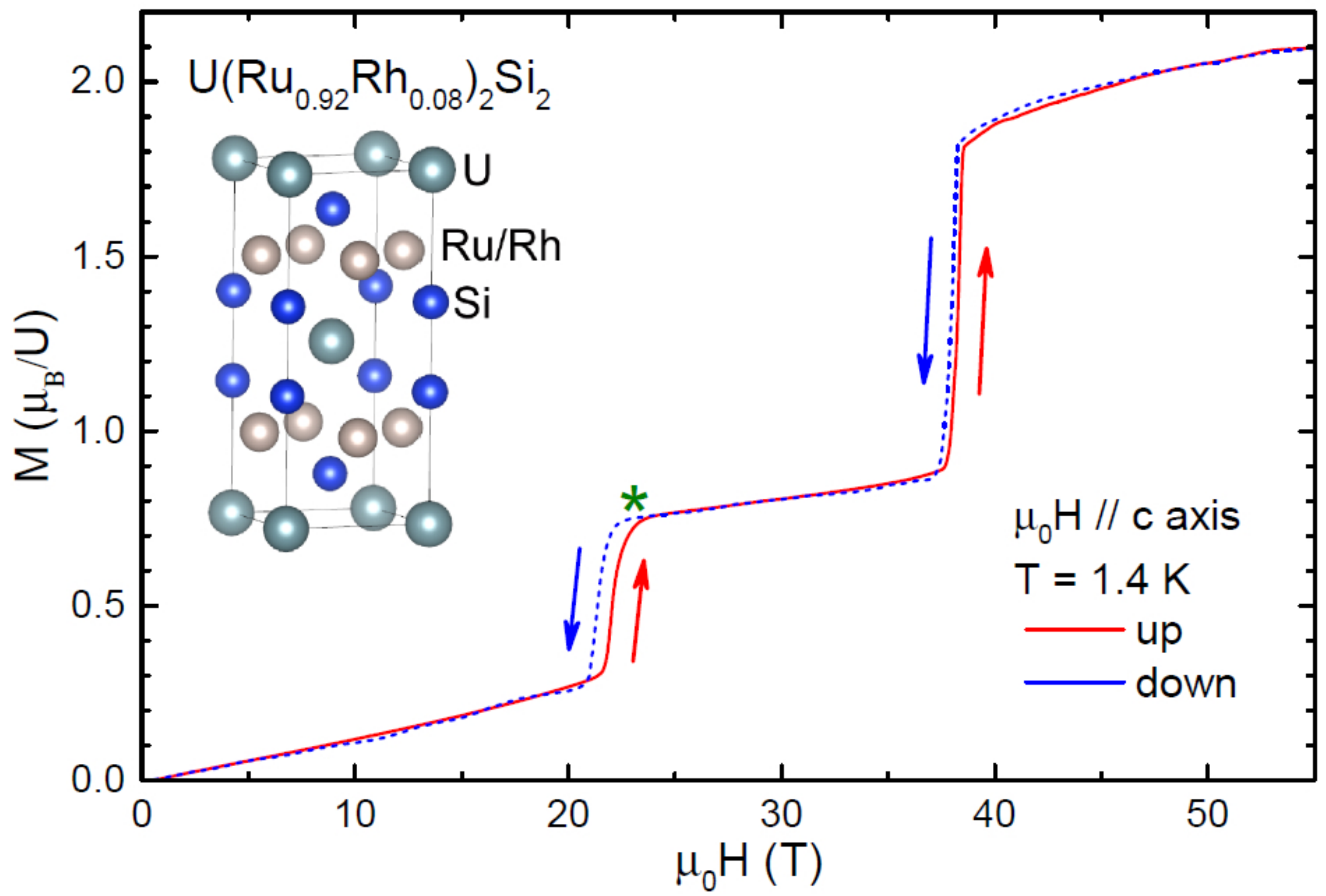}
\caption{(Color online) The field dependence of the magnetization of \urxrs~measured with field applied along the $c$ axis at 1.4 K. The star marks the field at which the neutron experiment has been performed. In the inset the tetragonal crystal structure adopted by the system is shown.} \label{fig1}
\label{fig1}
\end{figure}

Physical properties of \urxrs~are closely related to those of \urs~that include the heavy-fermion behavior but show the absence of both, the HO and SC states. The system does not order magnetically, only short-range order (SRO) characterized by \textbf{\textit{Q$_ {3}$}} at low temperatures is detected \cite{urrshfn16}. For fields up to 58 T applied along the $a$ axis (not shown) the magnetization is tiny and increases linearly with field. Oppositely, the $c$ axis magnetization exhibits a dramatic step-like increase at 22 T (Fig.~\ref{fig1}) followed by another one around 38 T. On decreasing the field a small hysteresis is seen at lower transition leading to the average critical field of $\mu_{0}$H$_{c1}$ = 21.6 T, the lowest critical field among Rh-doped \urrs~\cite{urrshfn28}. In the high field limit the magnetization tends to saturate at a level of 2.1 $\mu_{B}$/U. The magnetization step across the first MT amounts to 0.46 $\mu_{B}$/U and across the second MT at 38 T to 0.94 $\mu_{B}$/U. The increase at the former MT amounts to one third of the magnetization increase due to both MTs. 

\begin{figure}
\includegraphics*[scale=0.3]{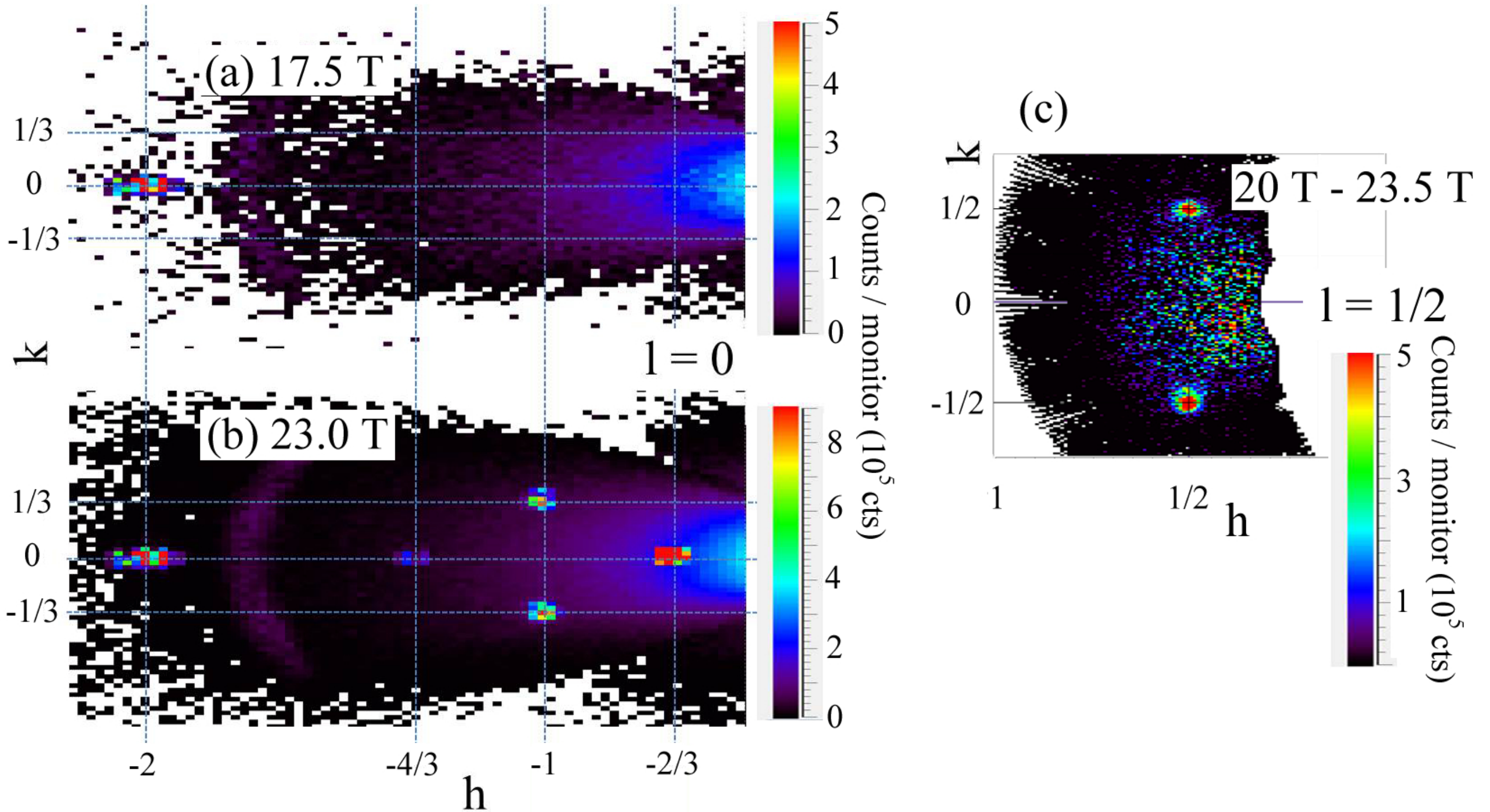}
\caption{(Color online) (a) Portion of the $l$ = 0 reciprocal space plane of \urrs~covered by the first geometry with field along the $c$ axis in the forward direction, showing the color-coded intensity of diffracted neutrons at 17.5 T, and (b) at 23 T, both at 1.4 K. (c) The intensity difference for the $l$ = $\frac{1}{2}$ reciprocal space plane recorded with field of 20 T and 23.5 T in the second geometry with field inclined from the $c$ axis.} 
\label{fig2}
\end{figure}

\begin{figure}
\includegraphics*[scale=0.32]{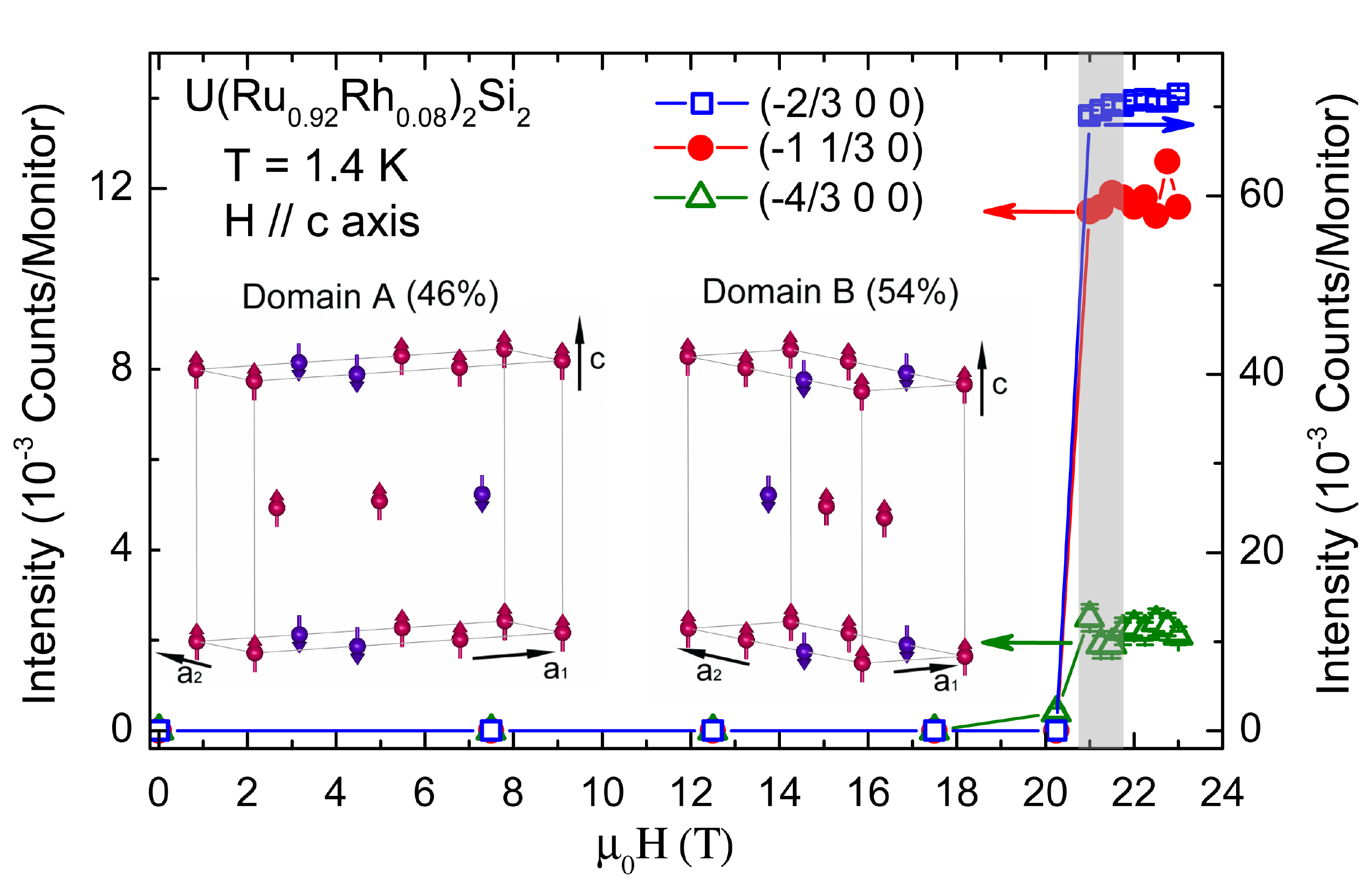}
\caption{(Color online) Field dependence of the intensity of representative magnetic reflections observed in \urxrs~single crystal at 1.4 K with the field applied along the $c$ axis. In the inset we show the field-induced magnetic structure consisting of two single-\textbf{\textit{Q}} domains. The shaded area denote the range of fields where the MT with lowering the field occurs.} 
\label{fig3}
\end{figure}

In Fig.~\ref{fig2} (a) and (b) we show a portion of the diffracted intensity distribution in the reciprocal space detected in the forward direction in 17.5 T and 23 T directed along the $c$ axis, respectively. In addition to the 200 nuclear reflection visible at both fields, new Bragg reflections having fractionalized indices are visible at 23 T. These reflections, that are resolution limited and are of magnetic origin, can be indexed by two, (in tetragonal symmetry equivalent) propagation vectors \textbf{\textit{Q$^{A}_ {2}$}} = ($\frac{2}{3}$ 0 0) and \textbf{\textit{Q$^{B}_ {2}$}} = (0 $\frac{2}{3}$ 0) and their associated \textbf{\textit{-Q$^{A}_ {2}$}} and \textbf{\textit{-Q$^{B}_ {2}$}} vectors, respectively. As shown in Fig.~\ref{fig3}, the intensities integrated around these positions show a step-like increase at MT, in agreement with the magnetization. The data were collected with decreasing fields after the sample being exposed to 24 T. The propagation vector, that does not change above the MT, suggests that the field-induced phase is commensurate with crystal lattice. No diffracted intensity has been observed at $\bar{1}$00, ($\bar{\frac{5}{3}}$ 0 0), ($\bar{\frac{2}{3}}$ $\frac{1}{3}$ 0) and similar reciprocal positions (see Fig.~\ref{fig2} (b)).

To resolve the mutual coupling between U moments of this phase we have used the representation analysis \cite{urrshfn26} that resulted in very few possible configurations (see Supplemental Material \cite{urrshfn31}). The best agreement is found for up-up-down sequence of U moments propagating along the $a$ axis. There are two single-\textbf{\textit{Q}} domains (domain A with \textbf{\textit{Q$^{A}_ {2}$}} = ($\frac{2}{3}$ 0 0) and domain B with \textbf{\textit{Q$^{B}_ {2}$}} = (0 $\frac{2}{3}$ 0)). Assuming that U moments order in both domains with identical moment magnitudes, the best fit leads to 1.45 (9) $\mu_{B}$/U and population of 46(1) and 54(1) vol.\% as shown in the inset of Fig.~\ref{fig3}. Such an arrangement implies a tendency towards crystal structure distortion because of ferromagnetic and AF couplings along originally equivalent $a$ axes, which is, however, not realized. We note that another, so called double-\textbf{\textit{Q}} structure yielding the same agreement with observed intensities cannot be in principle excluded. However, this structure is unlikely as it consists from highly unequal magnetic moments residing on equivalent crystallographic sites.

\begin{figure}
\includegraphics*[scale=0.39]{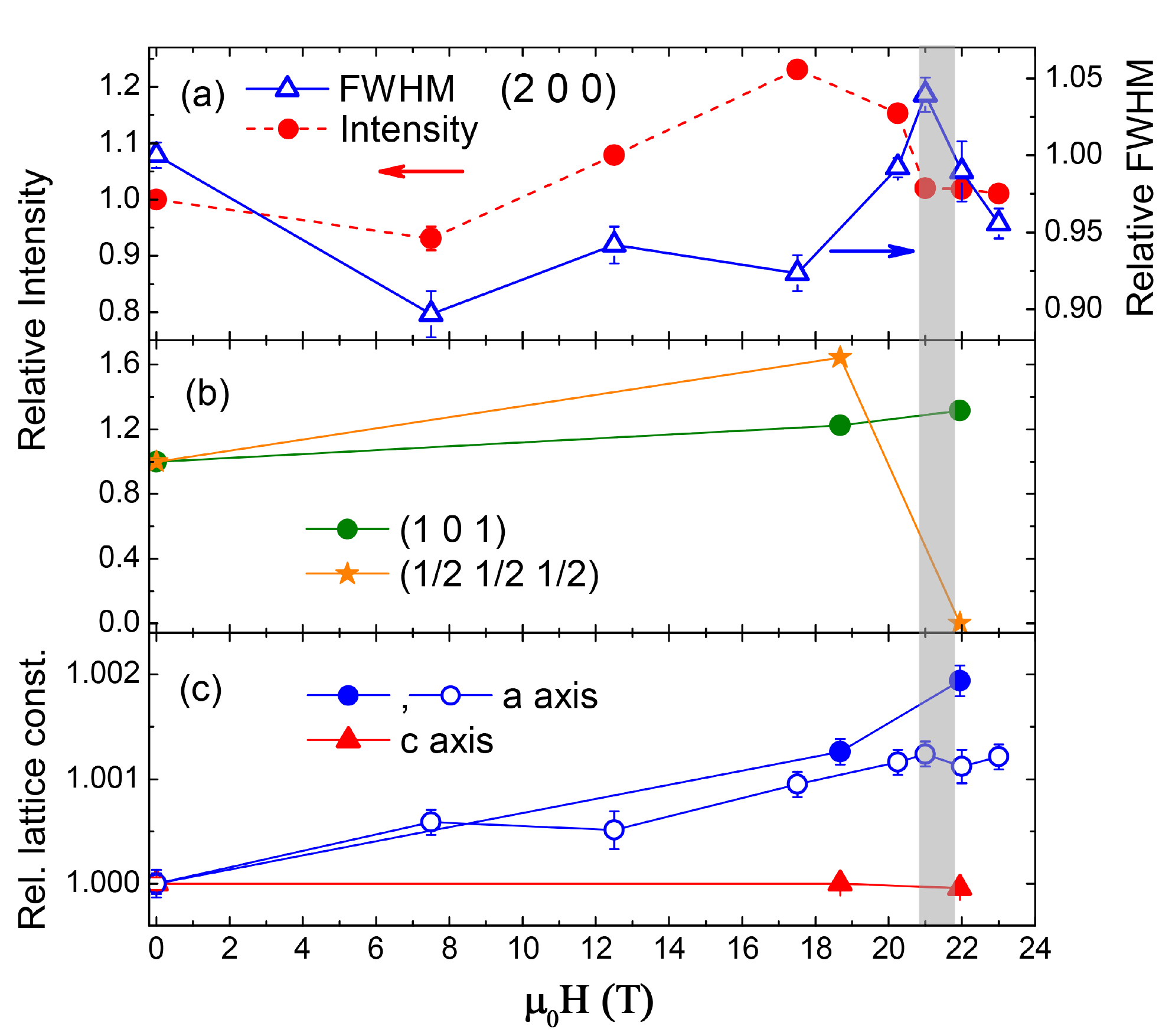}
\caption{(Color online) (a) Field dependence of the relative intensity of the 200 Bragg reflection along with its FWHM. (b) Field dependence of the relative intensity of the 101 Bragg reflection and of the SRO signal at ($\frac{1}{2}$ $\frac{1}{2}$ $\frac{1}{2}$). (c) Field dependence of the relative lattice constants as determined from measurement with field along the $c$ axis (open points) and with the $c$ axis inclined by 21$^{\circ}$ (closed points). The field values with the $c$ axis inclined by 21$^{\circ}$ with respect to the applied field were recalculated to effective values projected along the $c$ axis.} 
\label{fig4}
\end{figure}

In Fig.~\ref{fig4} (a) we show the field dependence of the intensity of the 200 reflection, normalized to the zero field value after the application of the field. It increases by $\sim$ 20 \% just below $\mu_{0}$H$_{c1}$ = 21.6 T as compared to low and high-field values that are almost identical. As this reflection has a much stronger crystal structure contribution and is therefore almost insensitive to magnetic moment, we interpret this observation to be due to field-dependent extinction. Such a conclusion is corroborated by the field dependence of the 200 reflection's relative full width at half maximum (FWHM) as shown also in Fig.~\ref{fig4} (a) establishing relation with the intensity. Interestingly, the FWHM peaks around $\mu_{0}$H$_{c1}$, where the MT transition takes place but shows at zero field a distinctly larger value than at intermediate fields where also the intensity is somewhat lower than at zero field. These observations suggest a presence of a field-induced strain influencing the crystal quality. However, the anticipated crystal structure distortion above MT is not realized, otherwise the 200 reflection would split as in the case of putative symmetry lowering of \urs~\cite{urrshfn25}. Comparing the data obtained at 23 and 0 T, we do not observe crystallographic distortion within our precision at the level of $\sim$ 2$\cdot$10$^{-4}$. 

The refined up-up-down U moment arrangement leads to a net ferromagnetic moment in agreement with magnetization data (see Fig.~\ref{fig1}) and imposes an increase of relevant nuclear reflections intensities. As shown in Fig.~\ref{fig4} (b), the intensity of the 101 reflection increases at 21.94 T by 32 \%. This increase is notably larger than the expected increase due to 0.46 $\mu_{B}$/U seen at the first MT. Since the nuclear 101 reflection is very sensitive to position of Si atoms, a change of the only free positional parameter $z_{Si}$ by $\sim$ 0.01 r.l.u. from $z_{Si}$ = 0.37325 \cite{urrshfn16} to $\sim$ 0.363 induced by field could explain the observed increase. This mechanism has been named as a possible reason for the missing ferromagnetic signal on top of the 101 reflection in the pulsed experiment on \urs~\cite{urrshfn12}. However, we can discard such a scenario as the 107 reflection would need to change significantly its intensity as well. Therefore, we attribute 101 reflection intensity increase to changes in extinction and estimate the increase due to magnetic order to $\sim$ 8 \%.  

Although our crystal does not exhibit any of the phases present in the pure \urs, it does exhibit SRO characterized by \textbf{\textit{Q$_ {3}$}} = ($\frac{1}{2}$ $\frac{1}{2}$ $\frac{1}{2}$) \cite{urrshfn16}. Such a propagation vector imposes four spatially disjunct single-\textbf{\textit{Q}} domains. Two of the domains, represented by reflections ($\frac{1}{2}$ $\frac{1}{2}$ $\frac{1}{2}$) and ($\frac{1}{2}$ -$\frac{1}{2}$ $\frac{1}{2}$), could be recorded with field inclined by 21$^{\circ}$ from the $c$ axis (see Fig.~\ref{fig2} (c)). Although not immediately apparent from the figure, they are not resolution limited and about three times broader than other Bragg reflections. This is in agreement with our previous study \cite{urrshfn16}. In Fig.~\ref{fig4} (b) we show the field dependence of the intensity of one of these peaks at three representative fields. It is surprising that the intensity of these peaks initially increases with increasing field before they disappear above the MT. One could speculate that this is because of domain redistribution but the observation is supported by experiments in fields up to 14.5 T, which also show an increase of all SRO peaks even with much smaller inclination of field with respect to the $c$ axis \cite{urrshfn23}. The increase followed by a sudden disappearance of these reflections is accompanied by the simultaneous appearance of \textbf{\textit{Q$_ {2}$}} reflections above $\mu_{0}$H$_{c1}$. Further we note that the background signal does not change across the MT. 

These observations can be explained by the itinerant character of 5$f^{2}$(U$^{4+}$) states situated in the vicinity of the Fermi surface (FS) that have been shown to exist in the pure \urs~\cite{urrshfn21,urrshfn22,urrshfn27} and in 4  \% Rh doped system \cite{urrshfn29}. Here, the 5$f$ states are split, forming a narrow pseudogap over a portion of the Fermi surface. Although no $ab-initio$ calculations exist to date on our system it is expected that Rh-doping modifies the FS topology in favor of a long-range magnetic order characterized by \textbf{\textit{Q$_ {3}$}} that appear at higher Rh concentrations. The doping has apparently a tendency to stabilize 5$f$ sub-band split states. The application of strong fields reconstructs the FS and Zeeman splits the sub-bands further eventually leading to a phase with well developed U moments ordering with \textbf{\textit{Q$_ {2}$}}. 

For the pure \urs~\cite{urrshfn12} and 4 \% Rh-doped \cite{urrshfn10} systems, magnetic moments of $\sim$ 0.50(5) $\mu_{B}$/U and $\sim$ 0.6(1) $\mu_{B}$/U had been determined assuming equal magnetic domains population. However, these moment values agree with magnetization data only qualitatively. In the pure case the magnetic structure is SDW with no ferromagnetic component on the single available nuclear reflection. In the 4 \% system, the magnetic structure becomes commensurate with corresponding increase of the 110 nuclear Bragg reflection \cite{urrshfn10}. Both structures have been determined on a basis of two observable magnetic Bragg reflections. Our results based on six magnetic and three nuclear reflections are obtained at stable thermodynamical conditions lead to U moments of 1.45 (9) $\mu_{B}$ in agreement with magnetization data. At 23 T, the U moments are in agreement with magnetic form factor of the U$^{3+}$/U$^{4+}$ type.  

In Fig.~\ref{fig4} (c) we show the field dependence of the relative lattice constants determined from the $d$ spacings of most intense nuclear reflections. While the $a$ axis parameter increases by 0.1-0.2 \% at 23 T, the $c$ axis is field independent. This finding is to be compared with the field-induced sample length change along the $c$ axis in pure \urs~\cite{urrshfn17} that shrinks by $\sim$ 3$\cdot$10$^{-5}$ at the first MT. Surprisingly, the increase of the $a$ axis parameter is present already below the MT transition. This observation corroborates our working model of the itinerant character of U moments in \urxrs~and field-induced modifications of the FS leading to a stabilization of 5$f$ states, analogical to FS reconstruction in the pure system. \cite{urrshfn38,urrshfn39}
Although still controversial, it has been reported that the HO lifts the four-fold symmetry of the \urs~lattice already in zero field. \cite{urrshfn25,urrshfn7} In the case of our system it is evident that both, \textbf{\textit{Q$_ {3}$}} = ($\frac{1}{2}$ $\frac{1}{2}$ $\frac{1}{2}$) below $\mu_{0}$H$_{c1}$ and \textbf{\textit{Q$_ {2}$}} = ($\frac{2}{3}$ 0 0) phases above MT, brake the time reversal symmetry although the crystal structure remains within our error bars undistorted. It is remarkable that the magnetic structures in systems with and without the HO are so different, the first being a SDW and in the latter a commensurate one, consisting of two single-\textbf{\textit{Q}} domains. Further studies including $ab-initio$ calculations that take into account doping and magnetic fields are needed to disclose relation between HO and other types of order.

In conclusion, our study shows that above $\mu_{0}$H$_{c1}$ = 21.6 T an uncompensated antiferromagnetic structure defined by the propagation vector \textbf{\textit{Q$_ {2}$}} = ($\frac{2}{3}$ 0 0) exists in \urxrs. Two spatially disjunct single-\textbf{\textit{Q$_ {2}$}} domains are almost equally populated. U moments of 1.45 (9) $\mu_{B}$ directed along the $c$ axis are arranged in an up-up-down sequence propagating along the $a$ axis. However, further high-field studies using NMR or uniaxial pressure are necessary to discard definitely a possible double-\textbf{\textit{Q}} structure that would need to consists from very unequal moments between 0.4 and 2.4 $\mu_{B}$. The low field short-range order that emerges from the pure \urs~due to Rh-doping is initially strengthened by the field but disappears above MT. Our results could be explained by the itinerant model of magnetism. 

 We acknowledge the support of the HLD at HZDR, member of the European Magnetic Field Laboratory (EMFL). We also acknowledge A. de Visser form University of Amsterdam for help with the sample preparation and discussion with F. Bourdarot (CEA Grenoble) and F. Duc (LNCMI Toulouse).

\bibliography{urrs_HFN}

\begin{thebibliography}{30}%
\makeatletter
\providecommand \@ifxundefined [1]{%
 \@ifx{#1\undefined}
}%
\providecommand \@ifnum [1]{%
 \ifnum #1\expandafter \@firstoftwo
 \else \expandafter \@secondoftwo
 \fi
}%
\providecommand \@ifx [1]{%
 \ifx #1\expandafter \@firstoftwo
 \else \expandafter \@secondoftwo
 \fi
}%
\providecommand \natexlab [1]{#1}%
\providecommand \enquote  [1]{``#1''}%
\providecommand \bibnamefont  [1]{#1}%
\providecommand \bibfnamefont [1]{#1}%
\providecommand \citenamefont [1]{#1}%
\providecommand \href@noop [0]{\@secondoftwo}%
\providecommand \href [0]{\begingroup \@sanitize@url \@href}%
\providecommand \@href[1]{\@@startlink{#1}\@@href}%
\providecommand \@@href[1]{\endgroup#1\@@endlink}%
\providecommand \@sanitize@url [0]{\catcode `\\12\catcode `\$12\catcode
  `\&12\catcode `\#12\catcode `\^12\catcode `\_12\catcode `\%12\relax}%
\providecommand \@@startlink[1]{}%
\providecommand \@@endlink[0]{}%
\providecommand \url  [0]{\begingroup\@sanitize@url \@url }%
\providecommand \@url [1]{\endgroup\@href {#1}{\urlprefix }}%
\providecommand \urlprefix  [0]{URL }%
\providecommand \Eprint [0]{\href }%
\providecommand \doibase [0]{http://dx.doi.org/}%
\providecommand \selectlanguage [0]{\@gobble}%
\providecommand \bibinfo  [0]{\@secondoftwo}%
\providecommand \bibfield  [0]{\@secondoftwo}%
\providecommand \translation [1]{[#1]}%
\providecommand \BibitemOpen [0]{}%
\providecommand \bibitemStop [0]{}%
\providecommand \bibitemNoStop [0]{.\EOS\space}%
\providecommand \EOS [0]{\spacefactor3000\relax}%
\providecommand \BibitemShut  [1]{\csname bibitem#1\endcsname}%
\let\auto@bib@innerbib\@empty
\bibitem [{\citenamefont {Mydosh}\ and\ \citenamefont
  {Oppeneer}(2011)}]{urrshfn6}%
  \BibitemOpen
  \bibfield  {author} {\bibinfo {author} {\bibfnamefont {J.~A.}\ \bibnamefont
  {Mydosh}}\ and\ \bibinfo {author} {\bibfnamefont {P.~M.}\ \bibnamefont
  {Oppeneer}},\ }\href {\doibase 10.1103/RevModPhys.83.1301} {\bibfield
  {journal} {\bibinfo  {journal} {Rev. Mod. Phys.}\ }\textbf {\bibinfo {volume}
  {83}},\ \bibinfo {pages} {1301} (\bibinfo {year} {2011})}\BibitemShut
  {NoStop}%
\bibitem [{\citenamefont {Mydosh}\ and\ \citenamefont
  {Oppeneer}(2014)}]{urrshfn7}%
  \BibitemOpen
  \bibfield  {author} {\bibinfo {author} {\bibfnamefont {J.~A.}\ \bibnamefont
  {Mydosh}}\ and\ \bibinfo {author} {\bibfnamefont {P.~M.}\ \bibnamefont
  {Oppeneer}},\ }\href {\doibase 10.1080/14786435.2014.916428} {\bibfield
  {journal} {\bibinfo  {journal} {Phil. Mag.}\ }\textbf {\bibinfo {volume}
  {94}},\ \bibinfo {pages} {3642} (\bibinfo {year} {2014})}\BibitemShut
  {NoStop}%
\bibitem [{\citenamefont {Palstra}\ \emph {et~al.}(1985)\citenamefont
  {Palstra}, \citenamefont {Menovsky}, \citenamefont {Berg}, \citenamefont
  {Dirkmaat}, \citenamefont {Kes}, \citenamefont {Nieuwenhuys},\ and\
  \citenamefont {Mydosh}}]{urrshfn1}%
  \BibitemOpen
  \bibfield  {author} {\bibinfo {author} {\bibfnamefont {T.~T.~M.}\
  \bibnamefont {Palstra}}, \bibinfo {author} {\bibfnamefont {A.~A.}\
  \bibnamefont {Menovsky}}, \bibinfo {author} {\bibfnamefont {J.~v.~d.}\
  \bibnamefont {Berg}}, \bibinfo {author} {\bibfnamefont {A.~J.}\ \bibnamefont
  {Dirkmaat}}, \bibinfo {author} {\bibfnamefont {P.~H.}\ \bibnamefont {Kes}},
  \bibinfo {author} {\bibfnamefont {G.~J.}\ \bibnamefont {Nieuwenhuys}}, \ and\
  \bibinfo {author} {\bibfnamefont {J.~A.}\ \bibnamefont {Mydosh}},\ }\href
  {\doibase 10.1103/PhysRevLett.55.2727} {\bibfield  {journal} {\bibinfo
  {journal} {Phys. Rev. Lett.}\ }\textbf {\bibinfo {volume} {55}},\ \bibinfo
  {pages} {2727} (\bibinfo {year} {1985})}\BibitemShut {NoStop}%
\bibitem [{\citenamefont {Hasselbach}\ \emph {et~al.}(1991)\citenamefont
  {Hasselbach}, \citenamefont {Lejay},\ and\ \citenamefont
  {Flouquet}}]{urrshfn2}%
  \BibitemOpen
  \bibfield  {author} {\bibinfo {author} {\bibfnamefont {K.}~\bibnamefont
  {Hasselbach}}, \bibinfo {author} {\bibfnamefont {P.}~\bibnamefont {Lejay}}, \
  and\ \bibinfo {author} {\bibfnamefont {J.}~\bibnamefont {Flouquet}},\ }\href
  {\doibase http://dx.doi.org/10.1016/0375-9601(91)90180-G} {\bibfield
  {journal} {\bibinfo  {journal} {Phys. Lett. A}\ }\textbf {\bibinfo {volume}
  {156}},\ \bibinfo {pages} {313 } (\bibinfo {year} {1991})}\BibitemShut
  {NoStop}%
\bibitem [{\citenamefont {Bourdarot}\ \emph {et~al.}(2014)\citenamefont
  {Bourdarot}, \citenamefont {Raymond},\ and\ \citenamefont
  {Regnault}}]{urrshfn3}%
  \BibitemOpen
  \bibfield  {author} {\bibinfo {author} {\bibfnamefont {F.}~\bibnamefont
  {Bourdarot}}, \bibinfo {author} {\bibfnamefont {S.}~\bibnamefont {Raymond}},
  \ and\ \bibinfo {author} {\bibfnamefont {L.-P.}\ \bibnamefont {Regnault}},\
  }\href {\doibase 10.1080/14786435.2014.935513} {\bibfield  {journal}
  {\bibinfo  {journal} {Phil. Mag.}\ }\textbf {\bibinfo {volume} {94}},\
  \bibinfo {pages} {3702} (\bibinfo {year} {2014})}\BibitemShut {NoStop}%
\bibitem [{\citenamefont {Wiebe}\ \emph {et~al.}(2007)\citenamefont {Wiebe},
  \citenamefont {Janik}, \citenamefont {MacDougall}, \citenamefont {Luke},
  \citenamefont {Garrett}, \citenamefont {Zhou}, \citenamefont {Jo},
  \citenamefont {Balicas}, \citenamefont {Qiu}, \citenamefont {Copley},
  \citenamefont {Yamani},\ and\ \citenamefont {Buyers}}]{urrshfn37}%
  \BibitemOpen
  \bibfield  {author} {\bibinfo {author} {\bibfnamefont {C.~R.}\ \bibnamefont
  {Wiebe}}, \bibinfo {author} {\bibfnamefont {J.~A.}\ \bibnamefont {Janik}},
  \bibinfo {author} {\bibfnamefont {G.~J.}\ \bibnamefont {MacDougall}},
  \bibinfo {author} {\bibfnamefont {G.~M.}\ \bibnamefont {Luke}}, \bibinfo
  {author} {\bibfnamefont {J.~D.}\ \bibnamefont {Garrett}}, \bibinfo {author}
  {\bibfnamefont {H.~D.}\ \bibnamefont {Zhou}}, \bibinfo {author}
  {\bibfnamefont {Y.~J.}\ \bibnamefont {Jo}}, \bibinfo {author} {\bibfnamefont
  {L.}~\bibnamefont {Balicas}}, \bibinfo {author} {\bibfnamefont
  {Y.}~\bibnamefont {Qiu}}, \bibinfo {author} {\bibfnamefont {J.~R.~D.}\
  \bibnamefont {Copley}}, \bibinfo {author} {\bibfnamefont {Z.}~\bibnamefont
  {Yamani}}, \ and\ \bibinfo {author} {\bibfnamefont {W.~J.~L.}\ \bibnamefont
  {Buyers}},\ }\href {\doibase 10.1038/nphys522} {\bibfield  {journal}
  {\bibinfo  {journal} {Nat. Phys.}\ }\textbf {\bibinfo {volume} {3}},\
  \bibinfo {pages} {96} (\bibinfo {year} {2007})}\BibitemShut {NoStop}%
\bibitem [{\citenamefont {Kim}\ \emph {et~al.}(2004)\citenamefont {Kim},
  \citenamefont {Harrison}, \citenamefont {Amitsuka}, \citenamefont {Jorge},
  \citenamefont {Jaime},\ and\ \citenamefont {Mydosh}}]{urrshfn5}%
  \BibitemOpen
  \bibfield  {author} {\bibinfo {author} {\bibfnamefont {K.~H.}\ \bibnamefont
  {Kim}}, \bibinfo {author} {\bibfnamefont {N.}~\bibnamefont {Harrison}},
  \bibinfo {author} {\bibfnamefont {H.}~\bibnamefont {Amitsuka}}, \bibinfo
  {author} {\bibfnamefont {G.~A.}\ \bibnamefont {Jorge}}, \bibinfo {author}
  {\bibfnamefont {M.}~\bibnamefont {Jaime}}, \ and\ \bibinfo {author}
  {\bibfnamefont {J.~A.}\ \bibnamefont {Mydosh}},\ }\href {\doibase
  10.1103/PhysRevLett.93.206402} {\bibfield  {journal} {\bibinfo  {journal}
  {Phys. Rev. Lett.}\ }\textbf {\bibinfo {volume} {93}},\ \bibinfo {pages}
  {206402} (\bibinfo {year} {2004})}\BibitemShut {NoStop}%
\bibitem [{\citenamefont {Kim}\ \emph {et~al.}(2003)\citenamefont {Kim},
  \citenamefont {Harrison}, \citenamefont {Jaime}, \citenamefont {Boebinger},\
  and\ \citenamefont {Mydosh}}]{urrshfn8}%
  \BibitemOpen
  \bibfield  {author} {\bibinfo {author} {\bibfnamefont {K.~H.}\ \bibnamefont
  {Kim}}, \bibinfo {author} {\bibfnamefont {N.}~\bibnamefont {Harrison}},
  \bibinfo {author} {\bibfnamefont {M.}~\bibnamefont {Jaime}}, \bibinfo
  {author} {\bibfnamefont {G.~S.}\ \bibnamefont {Boebinger}}, \ and\ \bibinfo
  {author} {\bibfnamefont {J.~A.}\ \bibnamefont {Mydosh}},\ }\href {\doibase
  10.1103/PhysRevLett.91.256401} {\bibfield  {journal} {\bibinfo  {journal}
  {Phys. Rev. Lett.}\ }\textbf {\bibinfo {volume} {91}},\ \bibinfo {pages}
  {256401} (\bibinfo {year} {2003})}\BibitemShut {NoStop}%
\bibitem [{\citenamefont {Burlet}\ \emph {et~al.}(1992)\citenamefont {Burlet},
  \citenamefont {Bourdarot}, \citenamefont {Quezel}, \citenamefont
  {Rossat-Mignod}, \citenamefont {Lejay}, \citenamefont {Chevalier},\ and\
  \citenamefont {Hickey}}]{urrshfn4}%
  \BibitemOpen
  \bibfield  {author} {\bibinfo {author} {\bibfnamefont {P.}~\bibnamefont
  {Burlet}}, \bibinfo {author} {\bibfnamefont {F.}~\bibnamefont {Bourdarot}},
  \bibinfo {author} {\bibfnamefont {S.}~\bibnamefont {Quezel}}, \bibinfo
  {author} {\bibfnamefont {J.}~\bibnamefont {Rossat-Mignod}}, \bibinfo {author}
  {\bibfnamefont {P.}~\bibnamefont {Lejay}}, \bibinfo {author} {\bibfnamefont
  {B.}~\bibnamefont {Chevalier}}, \ and\ \bibinfo {author} {\bibfnamefont
  {H.}~\bibnamefont {Hickey}},\ }\href {\doibase
  http://dx.doi.org/10.1016/0304-8853(92)91411-L} {\bibfield  {journal}
  {\bibinfo  {journal} {J. Magn. Magn. Mater.}\ }\textbf {\bibinfo {volume}
  {108}},\ \bibinfo {pages} {202 } (\bibinfo {year} {1992})}\BibitemShut
  {NoStop}%
\bibitem [{\citenamefont {Yokoyama}\ \emph {et~al.}(2004)\citenamefont
  {Yokoyama}, \citenamefont {Amitsuka}, \citenamefont {Itoh}, \citenamefont
  {Kawasaki}, \citenamefont {Tenya},\ and\ \citenamefont
  {Yoshizawa}}]{urrshfn32}%
  \BibitemOpen
  \bibfield  {author} {\bibinfo {author} {\bibfnamefont {M.}~\bibnamefont
  {Yokoyama}}, \bibinfo {author} {\bibfnamefont {H.}~\bibnamefont {Amitsuka}},
  \bibinfo {author} {\bibfnamefont {S.}~\bibnamefont {Itoh}}, \bibinfo {author}
  {\bibfnamefont {I.}~\bibnamefont {Kawasaki}}, \bibinfo {author}
  {\bibfnamefont {K.}~\bibnamefont {Tenya}}, \ and\ \bibinfo {author}
  {\bibfnamefont {H.}~\bibnamefont {Yoshizawa}},\ }\href {\doibase
  10.1143/JPSJ.73.545} {\bibfield  {journal} {\bibinfo  {journal} {J. Phys.
  Soc. Jap.}\ }\textbf {\bibinfo {volume} {73}},\ \bibinfo {pages} {545}
  (\bibinfo {year} {2004})}\BibitemShut {NoStop}%
\bibitem [{\citenamefont {Islam}\ \emph {et~al.}(2009)\citenamefont {Islam},
  \citenamefont {Ruff}, \citenamefont {Nojiri}, \citenamefont {Matsuda},
  \citenamefont {Ross}, \citenamefont {Gaulin}, \citenamefont {Qu},\ and\
  \citenamefont {Lang}}]{urrshfn9}%
  \BibitemOpen
  \bibfield  {author} {\bibinfo {author} {\bibfnamefont {Z.}~\bibnamefont
  {Islam}}, \bibinfo {author} {\bibfnamefont {J.~P.~C.}\ \bibnamefont {Ruff}},
  \bibinfo {author} {\bibfnamefont {H.}~\bibnamefont {Nojiri}}, \bibinfo
  {author} {\bibfnamefont {Y.~H.}\ \bibnamefont {Matsuda}}, \bibinfo {author}
  {\bibfnamefont {K.~A.}\ \bibnamefont {Ross}}, \bibinfo {author}
  {\bibfnamefont {B.~D.}\ \bibnamefont {Gaulin}}, \bibinfo {author}
  {\bibfnamefont {Z.}~\bibnamefont {Qu}}, \ and\ \bibinfo {author}
  {\bibfnamefont {J.~C.}\ \bibnamefont {Lang}},\ }\href {\doibase
  10.1063/1.3251273} {\bibfield  {journal} {\bibinfo  {journal} {Rev. Sci.
  Instr.}\ }\textbf {\bibinfo {volume} {80}},\ \bibinfo {pages} {113902}
  (\bibinfo {year} {2009})}\BibitemShut {NoStop}%
\bibitem [{\citenamefont {Knafo}\ \emph {et~al.}(2016)\citenamefont {Knafo},
  \citenamefont {Duc}, \citenamefont {Bourdarot}, \citenamefont {Kuwahara},
  \citenamefont {Nojiri}, \citenamefont {Aoki}, \citenamefont {Billette},
  \citenamefont {Frings}, \citenamefont {Tonon}, \citenamefont
  {Lelievre-Berna}, \citenamefont {Flouquet},\ and\ \citenamefont
  {Regnault}}]{urrshfn12}%
  \BibitemOpen
  \bibfield  {author} {\bibinfo {author} {\bibfnamefont {W.}~\bibnamefont
  {Knafo}}, \bibinfo {author} {\bibfnamefont {F.}~\bibnamefont {Duc}}, \bibinfo
  {author} {\bibfnamefont {F.}~\bibnamefont {Bourdarot}}, \bibinfo {author}
  {\bibfnamefont {K.}~\bibnamefont {Kuwahara}}, \bibinfo {author}
  {\bibfnamefont {H.}~\bibnamefont {Nojiri}}, \bibinfo {author} {\bibfnamefont
  {D.}~\bibnamefont {Aoki}}, \bibinfo {author} {\bibfnamefont {J.}~\bibnamefont
  {Billette}}, \bibinfo {author} {\bibfnamefont {P.}~\bibnamefont {Frings}},
  \bibinfo {author} {\bibfnamefont {X.}~\bibnamefont {Tonon}}, \bibinfo
  {author} {\bibfnamefont {E.}~\bibnamefont {Lelievre-Berna}}, \bibinfo
  {author} {\bibfnamefont {J.}~\bibnamefont {Flouquet}}, \ and\ \bibinfo
  {author} {\bibfnamefont {L.~P.}\ \bibnamefont {Regnault}},\ }\href@noop {}
  {\bibfield  {journal} {\bibinfo  {journal} {Nat. Comm.}\ }\textbf {\bibinfo
  {volume} {7}},\ \bibinfo {pages} {13075} (\bibinfo {year}
  {2016})}\BibitemShut {NoStop}%
\bibitem [{\citenamefont {Kuwahara}\ \emph {et~al.}(2013)\citenamefont
  {Kuwahara}, \citenamefont {Yoshii}, \citenamefont {Nojiri}, \citenamefont
  {Aoki}, \citenamefont {Knafo}, \citenamefont {Duc}, \citenamefont
  {Fabr\`eges}, \citenamefont {Scheerer}, \citenamefont {Frings}, \citenamefont
  {Rikken}, \citenamefont {Bourdarot}, \citenamefont {Regnault},\ and\
  \citenamefont {Flouquet}}]{urrshfn10}%
  \BibitemOpen
  \bibfield  {author} {\bibinfo {author} {\bibfnamefont {K.}~\bibnamefont
  {Kuwahara}}, \bibinfo {author} {\bibfnamefont {S.}~\bibnamefont {Yoshii}},
  \bibinfo {author} {\bibfnamefont {H.}~\bibnamefont {Nojiri}}, \bibinfo
  {author} {\bibfnamefont {D.}~\bibnamefont {Aoki}}, \bibinfo {author}
  {\bibfnamefont {W.}~\bibnamefont {Knafo}}, \bibinfo {author} {\bibfnamefont
  {F.}~\bibnamefont {Duc}}, \bibinfo {author} {\bibfnamefont {X.}~\bibnamefont
  {Fabr\`eges}}, \bibinfo {author} {\bibfnamefont {G.~W.}\ \bibnamefont
  {Scheerer}}, \bibinfo {author} {\bibfnamefont {P.}~\bibnamefont {Frings}},
  \bibinfo {author} {\bibfnamefont {G.~L. J.~A.}\ \bibnamefont {Rikken}},
  \bibinfo {author} {\bibfnamefont {F.}~\bibnamefont {Bourdarot}}, \bibinfo
  {author} {\bibfnamefont {L.~P.}\ \bibnamefont {Regnault}}, \ and\ \bibinfo
  {author} {\bibfnamefont {J.}~\bibnamefont {Flouquet}},\ }\href {\doibase
  10.1103/PhysRevLett.110.216406} {\bibfield  {journal} {\bibinfo  {journal}
  {Phys. Rev. Lett.}\ }\textbf {\bibinfo {volume} {110}},\ \bibinfo {pages}
  {216406} (\bibinfo {year} {2013})}\BibitemShut {NoStop}%
\bibitem [{\citenamefont {Proke\ifmmode~\check{s}\else \v{s}\fi{}}\ \emph
  {et~al.}(2017{\natexlab{a}})\citenamefont {Proke\ifmmode~\check{s}\else
  \v{s}\fi{}}, \citenamefont {Huang}, \citenamefont {Reehuis}, \citenamefont
  {Klemke}, \citenamefont {Hoffmann}, \citenamefont {Sokolowski}, \citenamefont
  {de~Visser},\ and\ \citenamefont {Mydosh}}]{urrshfn16}%
  \BibitemOpen
  \bibfield  {author} {\bibinfo {author} {\bibfnamefont {K.}~\bibnamefont
  {Proke\ifmmode~\check{s}\else \v{s}\fi{}}}, \bibinfo {author} {\bibfnamefont
  {Y.-K.}\ \bibnamefont {Huang}}, \bibinfo {author} {\bibfnamefont
  {M.}~\bibnamefont {Reehuis}}, \bibinfo {author} {\bibfnamefont
  {B.}~\bibnamefont {Klemke}}, \bibinfo {author} {\bibfnamefont {J.-U.}\
  \bibnamefont {Hoffmann}}, \bibinfo {author} {\bibfnamefont {A.}~\bibnamefont
  {Sokolowski}}, \bibinfo {author} {\bibfnamefont {A.}~\bibnamefont
  {de~Visser}}, \ and\ \bibinfo {author} {\bibfnamefont {J.~A.}\ \bibnamefont
  {Mydosh}},\ }\href {\doibase 10.1103/PhysRevB.95.035138} {\bibfield
  {journal} {\bibinfo  {journal} {Phys. Rev. B}\ }\textbf {\bibinfo {volume}
  {95}},\ \bibinfo {pages} {035138} (\bibinfo {year}
  {2017}{\natexlab{a}})}\BibitemShut {NoStop}%
\bibitem [{\citenamefont {Smeibidl}\ \emph {et~al.}(2010)\citenamefont
  {Smeibidl}, \citenamefont {Tennant}, \citenamefont {Ehmler},\ and\
  \citenamefont {Bird}}]{urrshfn13}%
  \BibitemOpen
  \bibfield  {author} {\bibinfo {author} {\bibfnamefont {P.}~\bibnamefont
  {Smeibidl}}, \bibinfo {author} {\bibfnamefont {A.}~\bibnamefont {Tennant}},
  \bibinfo {author} {\bibfnamefont {H.}~\bibnamefont {Ehmler}}, \ and\ \bibinfo
  {author} {\bibfnamefont {M.}~\bibnamefont {Bird}},\ }\href {\doibase
  10.1007/s10909-009-0062-1} {\bibfield  {journal} {\bibinfo  {journal} {J. Low
  Temp. Phys.}\ }\textbf {\bibinfo {volume} {159}},\ \bibinfo {pages} {402}
  (\bibinfo {year} {2010})}\BibitemShut {NoStop}%
\bibitem [{\citenamefont {Smeibidl}\ \emph {et~al.}(2016)\citenamefont
  {Smeibidl}, \citenamefont {Bird}, \citenamefont {Ehmler}, \citenamefont
  {Dixon}, \citenamefont {Heinrich}, \citenamefont {Hoffmann}, \citenamefont
  {Kempfer}, \citenamefont {Bole}, \citenamefont {Toth}, \citenamefont
  {Prokhnenko},\ and\ \citenamefont {Lake}}]{urrshfn18}%
  \BibitemOpen
  \bibfield  {author} {\bibinfo {author} {\bibfnamefont {P.}~\bibnamefont
  {Smeibidl}}, \bibinfo {author} {\bibfnamefont {M.}~\bibnamefont {Bird}},
  \bibinfo {author} {\bibfnamefont {H.}~\bibnamefont {Ehmler}}, \bibinfo
  {author} {\bibfnamefont {I.}~\bibnamefont {Dixon}}, \bibinfo {author}
  {\bibfnamefont {J.}~\bibnamefont {Heinrich}}, \bibinfo {author}
  {\bibfnamefont {M.}~\bibnamefont {Hoffmann}}, \bibinfo {author}
  {\bibfnamefont {S.}~\bibnamefont {Kempfer}}, \bibinfo {author} {\bibfnamefont
  {S.}~\bibnamefont {Bole}}, \bibinfo {author} {\bibfnamefont {J.}~\bibnamefont
  {Toth}}, \bibinfo {author} {\bibfnamefont {O.}~\bibnamefont {Prokhnenko}}, \
  and\ \bibinfo {author} {\bibfnamefont {B.}~\bibnamefont {Lake}},\ }\href
  {\doibase 10.1109/TASC.2016.2525773} {\bibfield  {journal} {\bibinfo
  {journal} {IEEE Trans. Appl. Supercond.}\ }\textbf {\bibinfo {volume} {26}},\
  \bibinfo {pages} {4301606} (\bibinfo {year} {2016})}\BibitemShut {NoStop}%
\bibitem [{\citenamefont {Prokhnenko}\ \emph {et~al.}(2015)\citenamefont
  {Prokhnenko}, \citenamefont {Stein}, \citenamefont {Bleif}, \citenamefont
  {Fromme}, \citenamefont {Bartkowiak},\ and\ \citenamefont
  {Wilpert}}]{urrshfn14}%
  \BibitemOpen
  \bibfield  {author} {\bibinfo {author} {\bibfnamefont {O.}~\bibnamefont
  {Prokhnenko}}, \bibinfo {author} {\bibfnamefont {W.-D.}\ \bibnamefont
  {Stein}}, \bibinfo {author} {\bibfnamefont {H.-J.}\ \bibnamefont {Bleif}},
  \bibinfo {author} {\bibfnamefont {M.}~\bibnamefont {Fromme}}, \bibinfo
  {author} {\bibfnamefont {M.}~\bibnamefont {Bartkowiak}}, \ and\ \bibinfo
  {author} {\bibfnamefont {T.}~\bibnamefont {Wilpert}},\ }\href {\doibase
  10.1063/1.4913656} {\bibfield  {journal} {\bibinfo  {journal} {Rev. Sci.
  Instr.}\ }\textbf {\bibinfo {volume} {86}},\ \bibinfo {pages} {033102}
  (\bibinfo {year} {2015})}\BibitemShut {NoStop}%
\bibitem [{\citenamefont {Proke\ifmmode~\check{s}\else \v{s}\fi{}}\ \emph
  {et~al.}()\citenamefont {Proke\ifmmode~\check{s}\else \v{s}\fi{}},
  \citenamefont {Bartkowiak}, \citenamefont {Rivin}, \citenamefont
  {Prokhnenko}, \citenamefont {F\"orster}, \citenamefont {Gerisher},
  \citenamefont {Wahle},\ and\ \citenamefont {Mydosh}}]{urrshfn31}%
  \BibitemOpen
  \bibfield  {author} {\bibinfo {author} {\bibfnamefont {K.}~\bibnamefont
  {Proke\ifmmode~\check{s}\else \v{s}\fi{}}}, \bibinfo {author} {\bibfnamefont
  {M.}~\bibnamefont {Bartkowiak}}, \bibinfo {author} {\bibfnamefont
  {O.}~\bibnamefont {Rivin}}, \bibinfo {author} {\bibfnamefont
  {O.}~\bibnamefont {Prokhnenko}}, \bibinfo {author} {\bibfnamefont
  {T.}~\bibnamefont {F\"orster}}, \bibinfo {author} {\bibfnamefont
  {S.}~\bibnamefont {Gerisher}}, \bibinfo {author} {\bibfnamefont {Y.-K.}\
  \bibnamefont {Wahle}, \bibfnamefont {R.~Huang}}, \ and\ \bibinfo {author}
  {\bibfnamefont {J.~A.}\ \bibnamefont {Mydosh}},\ }\href {\doibase See
  Supplemental Material at} {\ See Supplemental Material at}\BibitemShut
  {NoStop}%
\bibitem [{\citenamefont {Scheerer}\ \emph {et~al.}(2012)\citenamefont
  {Scheerer}, \citenamefont {Knafo}, \citenamefont {Aoki},\ and\ \citenamefont
  {Flouquet}}]{urrshfn20}%
  \BibitemOpen
  \bibfield  {author} {\bibinfo {author} {\bibfnamefont {G.~W.}\ \bibnamefont
  {Scheerer}}, \bibinfo {author} {\bibfnamefont {W.}~\bibnamefont {Knafo}},
  \bibinfo {author} {\bibfnamefont {D.}~\bibnamefont {Aoki}}, \ and\ \bibinfo
  {author} {\bibfnamefont {J.}~\bibnamefont {Flouquet}},\ }\href@noop {}
  {\bibfield  {journal} {\bibinfo  {journal} {J. Phys. Soc. Jap.}\ }\textbf
  {\bibinfo {volume} {81}},\ \bibinfo {pages} {SB005} (\bibinfo {year}
  {2012})}\BibitemShut {NoStop}%
\bibitem [{\citenamefont {Amitsuka}\ \emph {et~al.}(1990)\citenamefont
  {Amitsuka}, \citenamefont {Sakakibara},\ and\ \citenamefont
  {Miyako}}]{urrshfn28}%
  \BibitemOpen
  \bibfield  {author} {\bibinfo {author} {\bibfnamefont {H.}~\bibnamefont
  {Amitsuka}}, \bibinfo {author} {\bibfnamefont {T.}~\bibnamefont
  {Sakakibara}}, \ and\ \bibinfo {author} {\bibfnamefont {Y.}~\bibnamefont
  {Miyako}},\ }\href {\doibase http://dx.doi.org/10.1016/S0304-8853(10)80188-X}
  {\bibfield  {journal} {\bibinfo  {journal} {J. Mag. Mag. Mater.}\ }\textbf
  {\bibinfo {volume} {90}},\ \bibinfo {pages} {517 } (\bibinfo {year}
  {1990})}\BibitemShut {NoStop}%
\bibitem [{\citenamefont {Bertaut}(1981)}]{urrshfn26}%
  \BibitemOpen
  \bibfield  {author} {\bibinfo {author} {\bibfnamefont {E.}~\bibnamefont
  {Bertaut}},\ }\href {\doibase http://dx.doi.org/10.1016/0304-8853(81)90081-0}
  {\bibfield  {journal} {\bibinfo  {journal} {J. Mag. Mag. Mater.}\ }\textbf
  {\bibinfo {volume} {24}},\ \bibinfo {pages} {267 } (\bibinfo {year}
  {1981})}\BibitemShut {NoStop}%
\bibitem [{\citenamefont {Tonegawa}\ \emph {et~al.}(2014)\citenamefont
  {Tonegawa}, \citenamefont {Kasahara}, \citenamefont {Fukuda}, \citenamefont
  {Sugimoto}, \citenamefont {Yasuda}, \citenamefont {Tsuruhara}, \citenamefont
  {Watanabe}, \citenamefont {Mizukami}, \citenamefont {Haga}, \citenamefont
  {Matsuda}, \citenamefont {Yamamoto}, \citenamefont {Onuki}, \citenamefont
  {Ikeda}, \citenamefont {Matsuda},\ and\ \citenamefont
  {Shibauchi}}]{urrshfn25}%
  \BibitemOpen
  \bibfield  {author} {\bibinfo {author} {\bibfnamefont {S.}~\bibnamefont
  {Tonegawa}}, \bibinfo {author} {\bibfnamefont {S.}~\bibnamefont {Kasahara}},
  \bibinfo {author} {\bibfnamefont {T.}~\bibnamefont {Fukuda}}, \bibinfo
  {author} {\bibfnamefont {K.}~\bibnamefont {Sugimoto}}, \bibinfo {author}
  {\bibfnamefont {N.}~\bibnamefont {Yasuda}}, \bibinfo {author} {\bibfnamefont
  {Y.}~\bibnamefont {Tsuruhara}}, \bibinfo {author} {\bibfnamefont
  {D.}~\bibnamefont {Watanabe}}, \bibinfo {author} {\bibfnamefont
  {Y.}~\bibnamefont {Mizukami}}, \bibinfo {author} {\bibfnamefont
  {Y.}~\bibnamefont {Haga}}, \bibinfo {author} {\bibfnamefont {T.~D.}\
  \bibnamefont {Matsuda}}, \bibinfo {author} {\bibfnamefont {E.}~\bibnamefont
  {Yamamoto}}, \bibinfo {author} {\bibfnamefont {Y.}~\bibnamefont {Onuki}},
  \bibinfo {author} {\bibfnamefont {H.}~\bibnamefont {Ikeda}}, \bibinfo
  {author} {\bibfnamefont {Y.}~\bibnamefont {Matsuda}}, \ and\ \bibinfo
  {author} {\bibfnamefont {T.}~\bibnamefont {Shibauchi}},\ }\href@noop {}
  {\bibfield  {journal} {\bibinfo  {journal} {Nat. Com.}\ }\textbf {\bibinfo
  {volume} {5}},\ \bibinfo {pages} {4188} (\bibinfo {year} {2014})}\BibitemShut
  {NoStop}%
\bibitem [{\citenamefont {Proke\ifmmode~\check{s}\else \v{s}\fi{}}\ \emph
  {et~al.}(2017{\natexlab{b}})\citenamefont {Proke\ifmmode~\check{s}\else
  \v{s}\fi{}}, \citenamefont {Bartkowiak}, \citenamefont {Prokhnenko},
  \citenamefont {Rivin}, \citenamefont {Huang},\ and\ \citenamefont
  {Mydosh}}]{urrshfn23}%
  \BibitemOpen
  \bibfield  {author} {\bibinfo {author} {\bibfnamefont {K.}~\bibnamefont
  {Proke\ifmmode~\check{s}\else \v{s}\fi{}}}, \bibinfo {author} {\bibfnamefont
  {M.}~\bibnamefont {Bartkowiak}}, \bibinfo {author} {\bibfnamefont
  {O.}~\bibnamefont {Prokhnenko}}, \bibinfo {author} {\bibfnamefont
  {O.}~\bibnamefont {Rivin}}, \bibinfo {author} {\bibfnamefont {Y.-K.}\
  \bibnamefont {Huang}}, \ and\ \bibinfo {author} {\bibfnamefont {J.~A.}\
  \bibnamefont {Mydosh}},\ }\href@noop {} {\bibfield  {journal} {\bibinfo
  {journal} {to be published}\ } (\bibinfo {year}
  {2017}{\natexlab{b}})}\BibitemShut {NoStop}%
\bibitem [{\citenamefont {Oppeneer}\ \emph {et~al.}(2010)\citenamefont
  {Oppeneer}, \citenamefont {Rusz}, \citenamefont {Elgazzar}, \citenamefont
  {Suzuki}, \citenamefont {Durakiewicz},\ and\ \citenamefont
  {Mydosh}}]{urrshfn21}%
  \BibitemOpen
  \bibfield  {author} {\bibinfo {author} {\bibfnamefont {P.~M.}\ \bibnamefont
  {Oppeneer}}, \bibinfo {author} {\bibfnamefont {J.}~\bibnamefont {Rusz}},
  \bibinfo {author} {\bibfnamefont {S.}~\bibnamefont {Elgazzar}}, \bibinfo
  {author} {\bibfnamefont {M.-T.}\ \bibnamefont {Suzuki}}, \bibinfo {author}
  {\bibfnamefont {T.}~\bibnamefont {Durakiewicz}}, \ and\ \bibinfo {author}
  {\bibfnamefont {J.~A.}\ \bibnamefont {Mydosh}},\ }\href {\doibase
  10.1103/PhysRevB.82.205103} {\bibfield  {journal} {\bibinfo  {journal} {Phys.
  Rev. B}\ }\textbf {\bibinfo {volume} {82}},\ \bibinfo {pages} {205103}
  (\bibinfo {year} {2010})}\BibitemShut {NoStop}%
\bibitem [{\citenamefont {Haule}\ and\ \citenamefont
  {Kotliar}(2009)}]{urrshfn22}%
  \BibitemOpen
  \bibfield  {author} {\bibinfo {author} {\bibfnamefont {K.}~\bibnamefont
  {Haule}}\ and\ \bibinfo {author} {\bibfnamefont {G.}~\bibnamefont
  {Kotliar}},\ }\href {\doibase 10.1038/NPHYS1392} {\bibfield  {journal}
  {\bibinfo  {journal} {Nat. Phys.}\ }\textbf {\bibinfo {volume} {5}},\
  \bibinfo {pages} {796} (\bibinfo {year} {2009})}\BibitemShut {NoStop}%
\bibitem [{\citenamefont {Sakai}\ \emph {et~al.}(2014)\citenamefont {Sakai},
  \citenamefont {Tokunaga}, \citenamefont {Kambe}, \citenamefont {Urbano},
  \citenamefont {Suzuki}, \citenamefont {Kuhns}, \citenamefont {Reyes},
  \citenamefont {Tobash}, \citenamefont {Ronning}, \citenamefont {Bauer},\ and\
  \citenamefont {Thompson}}]{urrshfn27}%
  \BibitemOpen
  \bibfield  {author} {\bibinfo {author} {\bibfnamefont {H.}~\bibnamefont
  {Sakai}}, \bibinfo {author} {\bibfnamefont {Y.}~\bibnamefont {Tokunaga}},
  \bibinfo {author} {\bibfnamefont {S.}~\bibnamefont {Kambe}}, \bibinfo
  {author} {\bibfnamefont {R.~R.}\ \bibnamefont {Urbano}}, \bibinfo {author}
  {\bibfnamefont {M.-T.}\ \bibnamefont {Suzuki}}, \bibinfo {author}
  {\bibfnamefont {P.~L.}\ \bibnamefont {Kuhns}}, \bibinfo {author}
  {\bibfnamefont {A.~P.}\ \bibnamefont {Reyes}}, \bibinfo {author}
  {\bibfnamefont {P.~H.}\ \bibnamefont {Tobash}}, \bibinfo {author}
  {\bibfnamefont {F.}~\bibnamefont {Ronning}}, \bibinfo {author} {\bibfnamefont
  {E.~D.}\ \bibnamefont {Bauer}}, \ and\ \bibinfo {author} {\bibfnamefont
  {J.~D.}\ \bibnamefont {Thompson}},\ }\href {\doibase
  10.1103/PhysRevLett.112.236401} {\bibfield  {journal} {\bibinfo  {journal}
  {Phys. Rev. Lett.}\ }\textbf {\bibinfo {volume} {112}},\ \bibinfo {pages}
  {236401} (\bibinfo {year} {2014})}\BibitemShut {NoStop}%
\bibitem [{\citenamefont {Oh}\ \emph {et~al.}(2007)\citenamefont {Oh},
  \citenamefont {Kim}, \citenamefont {Sharma}, \citenamefont {Harrison},
  \citenamefont {Amitsuka},\ and\ \citenamefont {Mydosh}}]{urrshfn29}%
  \BibitemOpen
  \bibfield  {author} {\bibinfo {author} {\bibfnamefont {Y.~S.}\ \bibnamefont
  {Oh}}, \bibinfo {author} {\bibfnamefont {K.~H.}\ \bibnamefont {Kim}},
  \bibinfo {author} {\bibfnamefont {P.~A.}\ \bibnamefont {Sharma}}, \bibinfo
  {author} {\bibfnamefont {N.}~\bibnamefont {Harrison}}, \bibinfo {author}
  {\bibfnamefont {H.}~\bibnamefont {Amitsuka}}, \ and\ \bibinfo {author}
  {\bibfnamefont {J.~A.}\ \bibnamefont {Mydosh}},\ }\href {\doibase
  10.1103/PhysRevLett.98.016401} {\bibfield  {journal} {\bibinfo  {journal}
  {Phys. Rev. Lett.}\ }\textbf {\bibinfo {volume} {98}},\ \bibinfo {pages}
  {016401} (\bibinfo {year} {2007})}\BibitemShut {NoStop}%
\bibitem [{\citenamefont {Correa}\ \emph {et~al.}(2012)\citenamefont {Correa},
  \citenamefont {Francoual}, \citenamefont {Jaime}, \citenamefont {Harrison},
  \citenamefont {Murphy}, \citenamefont {Palm}, \citenamefont {Tozer},
  \citenamefont {Lacerda}, \citenamefont {Sharma},\ and\ \citenamefont
  {Mydosh}}]{urrshfn17}%
  \BibitemOpen
  \bibfield  {author} {\bibinfo {author} {\bibfnamefont {V.~F.}\ \bibnamefont
  {Correa}}, \bibinfo {author} {\bibfnamefont {S.}~\bibnamefont {Francoual}},
  \bibinfo {author} {\bibfnamefont {M.}~\bibnamefont {Jaime}}, \bibinfo
  {author} {\bibfnamefont {N.}~\bibnamefont {Harrison}}, \bibinfo {author}
  {\bibfnamefont {T.~P.}\ \bibnamefont {Murphy}}, \bibinfo {author}
  {\bibfnamefont {E.~C.}\ \bibnamefont {Palm}}, \bibinfo {author}
  {\bibfnamefont {S.~W.}\ \bibnamefont {Tozer}}, \bibinfo {author}
  {\bibfnamefont {A.~H.}\ \bibnamefont {Lacerda}}, \bibinfo {author}
  {\bibfnamefont {P.~A.}\ \bibnamefont {Sharma}}, \ and\ \bibinfo {author}
  {\bibfnamefont {J.~A.}\ \bibnamefont {Mydosh}},\ }\href {\doibase
  10.1103/PhysRevLett.109.246405} {\bibfield  {journal} {\bibinfo  {journal}
  {Phys. Rev. Lett.}\ }\textbf {\bibinfo {volume} {109}},\ \bibinfo {pages}
  {246405} (\bibinfo {year} {2012})}\BibitemShut {NoStop}%
\bibitem [{\citenamefont {Altarawneh}\ \emph {et~al.}(2011)\citenamefont
  {Altarawneh}, \citenamefont {Harrison}, \citenamefont {Sebastian},
  \citenamefont {Balicas}, \citenamefont {Tobash}, \citenamefont {Thompson},
  \citenamefont {Ronning},\ and\ \citenamefont {Bauer}}]{urrshfn38}%
  \BibitemOpen
  \bibfield  {author} {\bibinfo {author} {\bibfnamefont {M.~M.}\ \bibnamefont
  {Altarawneh}}, \bibinfo {author} {\bibfnamefont {N.}~\bibnamefont
  {Harrison}}, \bibinfo {author} {\bibfnamefont {S.~E.}\ \bibnamefont
  {Sebastian}}, \bibinfo {author} {\bibfnamefont {L.}~\bibnamefont {Balicas}},
  \bibinfo {author} {\bibfnamefont {P.~H.}\ \bibnamefont {Tobash}}, \bibinfo
  {author} {\bibfnamefont {J.~D.}\ \bibnamefont {Thompson}}, \bibinfo {author}
  {\bibfnamefont {F.}~\bibnamefont {Ronning}}, \ and\ \bibinfo {author}
  {\bibfnamefont {E.~D.}\ \bibnamefont {Bauer}},\ }\href {\doibase
  10.1103/PhysRevLett.106.146403} {\bibfield  {journal} {\bibinfo  {journal}
  {Phys. Rev. Lett.}\ }\textbf {\bibinfo {volume} {106}},\ \bibinfo {pages}
  {146403} (\bibinfo {year} {2011})}\BibitemShut {NoStop}%
\bibitem [{\citenamefont {Harrison}\ \emph {et~al.}(2013)\citenamefont
  {Harrison}, \citenamefont {Moll}, \citenamefont {Sebastian}, \citenamefont
  {Balicas}, \citenamefont {Altarawneh}, \citenamefont {Zhu}, \citenamefont
  {Tobash}, \citenamefont {Ronning}, \citenamefont {Bauer},\ and\ \citenamefont
  {Batlogg}}]{urrshfn39}%
  \BibitemOpen
  \bibfield  {author} {\bibinfo {author} {\bibfnamefont {N.}~\bibnamefont
  {Harrison}}, \bibinfo {author} {\bibfnamefont {P.~J.~W.}\ \bibnamefont
  {Moll}}, \bibinfo {author} {\bibfnamefont {S.~E.}\ \bibnamefont {Sebastian}},
  \bibinfo {author} {\bibfnamefont {L.}~\bibnamefont {Balicas}}, \bibinfo
  {author} {\bibfnamefont {M.~M.}\ \bibnamefont {Altarawneh}}, \bibinfo
  {author} {\bibfnamefont {J.-X.}\ \bibnamefont {Zhu}}, \bibinfo {author}
  {\bibfnamefont {P.~H.}\ \bibnamefont {Tobash}}, \bibinfo {author}
  {\bibfnamefont {F.}~\bibnamefont {Ronning}}, \bibinfo {author} {\bibfnamefont
  {E.~D.}\ \bibnamefont {Bauer}}, \ and\ \bibinfo {author} {\bibfnamefont
  {B.}~\bibnamefont {Batlogg}},\ }\href {\doibase 10.1103/PhysRevB.88.241108}
  {\bibfield  {journal} {\bibinfo  {journal} {Phys. Rev. B}\ }\textbf {\bibinfo
  {volume} {88}},\ \bibinfo {pages} {241108} (\bibinfo {year}
  {2013})}\BibitemShut {NoStop}%
\end{thebibliography}%

\end{document}